\documentclass{aastex631}

\usepackage{amsmath}
\usepackage{CJKutf8}
\shorttitle{NT Colors}
\shortauthors{Markwardt et al.}
\graphicspath{{./}{figures/}}

\begin{document}

\title{Photometric Survey of Neptune's Trojan Asteroids I: The Color Distribution}

\author[0000-0002-2486-1118]{Larissa Markwardt}
\affiliation{Department of Physics, University of Michigan \\
450 Church Street \\
Ann Arbor, MI 48109-1107, USA}

\author[0000-0001-7737-6784]{Hsing~Wen~Lin (\begin{CJK*}{UTF8}{gbsn}
林省文\end{CJK*})}
\affiliation{Department of Physics, University of Michigan \\
450 Church Street \\
Ann Arbor, MI 48109-1107, USA}

\author[0000-0001-6942-2736]{David Gerdes}
\affiliation{Department of Astronomy, University of Michigan \\
1085 South University Avenue \\
Ann Arbor, MI 48109-1107, USA}
\affiliation{Department of Physics, University of Michigan \\
450 Church Street \\
Ann Arbor, MI 48109-1107, USA}

\author[0000-0002-8167-1767]{Fred C. Adams}
\affiliation{Department of Astronomy, University of Michigan \\
1085 South University Avenue \\
Ann Arbor, MI 48109-1107, USA}
\affiliation{Department of Physics, University of Michigan \\
450 Church Street \\
Ann Arbor, MI 48109-1107, USA}

\correspondingauthor{Larissa Markwardt}
\email{lmmarkwa@umich.edu}



\begin{abstract}

In 2018, Jewitt identified the ``The Trojan Color Conundrum", namely that Neptune's Trojan asteroids (NTs) had no ultra-red members, unlike the the nearby Kuiper Belt. Since then, numerous ultra-red NTs have been discovered, seemingly resolving this conundrum \citep{2019Icar..321..426L, Bolin2023}. However, it is still unclear whether or not the Kuiper Belt has a color distribution consistent with the NT population, as would be expected if it were the source population. In this work, we present a new photometric survey of 15 out of 31 NTs. We utilized the Sloan $g^{\prime}$$r^{\prime}$$i^{\prime}$$z^{\prime}$ filters on the IMACS f/4 instrument which is mounted on the 6.5m Baade telescope. In this survey, we identify four NTs as being ultra-red using a Principal Component Analysis (PCA). This result brings the ratio of red to ultra-red NTs to 7.75:1, more consistent with the corresponding Trans-Neptunian Object (TNO) ratio of 4-11:1. 
We also identify three targets as being blue (nearly Solar) in color. Such objects may be C-type surfaces, but we see more of these blue NTs than has been observed in the Kuiper Belt \citep{2018ApJ...855L..26S}. Finally, we show that there are hints of a color-absolute magnitude (H) correlation, with larger H (smaller sized, lower albedo) tending to be more red, but more data is needed to confirm this result. The origin of such a correlation remains an open question which will be addressed by future observations of the surface composition of these targets and their rotational properties. 

\end{abstract}



\section{Introduction} \label{sec:intro}
 
Trojan asteroids are planetary companions that reside in the asymmetric 1:1 mean-motion resonance of planets; these asteroids librate at the planet-Sun L4 and L5 Lagrange points, meaning that they have the same orbit as the planet but librate about a point 60$^{\circ}$ ahead of (L4) or behind (L5) the planet. Numerical simulations show that orbits of Trojan asteroids can be quite stable, on order the age of the Solar System \citep{2012MNRAS.426.3051C, 2016A&A...592A.146G, 2011MNRAS.412..537L}. Therefore, the stable members of these populations are likely relatively undisturbed remnants of our primordial planetary disk. The physical properties of these populations can thus give us a window into the early Solar System. 

However, Neptune's Trojan asteroids are not thought to have formed \textit{in-situ}. Rather, this population likely grew through capture of planetesimals during the epoch of planetary migration, during which the outer planets migrated from the location of their formation to their present day locations \citep{1984Icar...58..109F, 1993Natur.365..819M, 1995AJ....110..420M, 1999AJ....117.3041H}. Assuming Neptune migrated significantly in its early evolution, the Lagrange points must have also migrated with it \citep{2004Icar..167..347K} Therefore, the NT population can be used to constrain migratory models \citep{2016A&A...592A.146G, 2013ApJ...768...45N, 2018NatAs...2..878N, 2017AJ....153..127P}. Such migration would have occurred in the first several hundred Myr in the history of the Solar System, so while these objects may not have formed \textit{in-situ}, they still are remnants of the very early Solar System.

Such models show that primordial Jupiter Trojan populations do not survive this planetary migration, indicating they must have originated from elsewhere in the Solar System.  \citep{2015AJ....150..186R}. Similarly, since the dynamics of planetary migration likely dispersed any primordial NTs as well, from where did the current population of NTs originate? The most likely source is the nearby Kuiper Belt. If that were the case, one would expect these two populations to be similar in size and color (surface composition). Regarding the color of the KBOs, the bimodality of red ($g-i$ \textless\ 1.2) vs. ultra-red ($g-i$ \textgreater\ 1.2) members has been well established \citep{2010AJ....139.1394S, 2011ApJS..197...31S, 2012A&A...546A.115H, 2012A&A...546A..86P, 2012AJ....144..169S, 2014ApJ...793L...2L, 2015A&A...577A..35P, 2017AJ....154..101P, 2017AJ....153..145W, 2019ApJS..243...12S}. Similarly, the centaur population, small bodies which orbit between Jupiter and Neptune, are thought to be fed by planetesimals escaping the NT region \citep{2010MNRAS.402...13H}. These objects are also red/ultra-red in color \citep{2012A&A...546A..86P, 2015A&A...577A..35P}. 

Through 2018, no ultra-red NTs had been found, making their color distribution distinctly different than their expected origins or offshoots. Termed the ``Trojan Color Conundrum", this tension is not easy to resolve \citep{2018AJ....155...56J}. One explanation is that some sort of resurfacing has happened to the NT population specifically that affected neither the centaurs or KBOs. Jupiter's Trojan population is also devoid of ultra-red members which is thought to be due to thermal resurfacing \citep{1996AJ....112.2310L, 2002AJ....123.1039J}. However, the temperatures at the distance of Neptune are too cold for such a scenario to be valid \citep{2018AJ....155...56J}. Another potential explanation is collisional resurfacing, which could strip the ultra-red crust off of the surfaces of these bodies revealing a bluer surface underneath. One source of such collisions could be Plutinos, 3:2 resonators with Neptune,   
which have significant orbital overlap with the NT population \citep{2009A&A...508.1021A}. Such collisions are expected to occur when Plutinos have high libration amplitudes, high eccentricities, and low inclinations; therefore, we would expect the color distribution of NTs to be inclination-dependent as well, where high inclination NTs avoid these collisions and retain their ultra-red surfaces. Finally, this discrepancy could be due to a primordial boundary between red/ultra-red bodies that was subsequently mixed by Neptune's migration \citep{2014Natur.505..629D, 2019ApJ...875...30N}. Based on the exact nature of the epochs of radial mixing, mass removal, and planet migration, the resulting NT population could be devoid of ultra-red members while the Centaur population is not \citep{2019ApJ...875...30N}, but specific simulations of these two populations have not been conducted. This hypothesis has been supported by the discovery of two Trans-Neptunian Object (TNO)-like (red) objects all the way in the asteroid belt \citep{2021ApJ...916L...6H}.

In 2019, the first ultra-red NT, 2013~VX$_{30}$, was discovered \citep{2019Icar..321..426L}, and additional ultra-red NTs have been discovered since then \citep{Bolin2023}. On the surface, these discoveries seem to resolve the conundrum. However, the color distribution of NTs still appears distinct from that of other TNO populations \citep{Bolin2023}. Further observations of the NT population are needed to determine whether or not these distributions are truly distinct.

The structure of this paper is as follows: Section 2 describes the design of our photometric survey. Section 3 outlines our data reduction process. Section 4 presents the results of our survey. Section 5 discuss the meaning of our results. Section 6 is conclusions drawn from these results. 

\section{Survey Design}\label{survey}
The goal of this paper is to measure the optical colors
of currently known NTs in order to better understand the physical characteristics of their surfaces. The targets are listed in Table \ref{tab:target_list}. All of our targets have been previously observed but not by the same survey. All of our targets, except 2015~VU$_{207}$, were already known to be stable for $\sim$Gyr \citep{2021Icar..36114391L, 2022RNAAS...6...79L}. Following the methods of \cite{2022RNAAS...6...79L}, we find that 2015~VU$_{207}$ is also stable for Gyr in our simulations. 

We used the IMACS f/4 instrument on the 6.5m Baade telescope at Las Campanas Observatory on 4 unique nights to observe this population. IMACS was most suitable for this task with its optical wavelength coverage ($\sim$400 - 900 nm) and large FOV to account for the positional uncertainty of the targets. The Sloan $g^{\prime}$$r^{\prime}$$i^{\prime}$$z^{\prime}$ filters were used for our photometric measurements. In order to account for any variation due to a target’s rotational period, we observed each target with ``bounding" $r^{\prime}$-band observations (i.e. each observation in a different filter was preceded and followed by an observation in $r^{\prime}$). . We chose $r^{\prime}$ to be the bounding observations since this filter reaches the highest SNR in the shortest amount of time. The fast readout mode with 2x2 binning was used. 

\begin{table}[h!]
    \centering
        \caption{NT targets of this survey. Columns: (1) Object Designation, has previous color measurements taken from 1: \cite{2012AJ....144..169S}, 2: \cite{2013AJ....145...96P}, 3: \cite{2018AJ....155...56J}, 4: \cite{2019Icar..321..426L}, 5: \cite{Bolin2023}; (2) Lagrange Point; (3) Eccentricity; (4) Inclination ($^{\circ}$); (5) Absolute Magnitude; (6) Dates observed; (7) Measured ave. SDSS r-band magnitude; (8) Measured SDSS g-r (mag); (9) Measured SDSS r-i (mag); (10) Measured SDSS r-z (mag); (11) Color classification determined based on the Principal Component Analysis (see Sec. 4.2).
}\resizebox{\textwidth}{!}{
    \begin{tabular}{lccccccccccc}
         Name & L4/L5 & e & i & H & Date Observed &  Ave. r & $g-r$ & $r-i$ & $r-z$  & Color Class.\\
         \hline \hline
         2006 RJ$_{103}$$^{1,2}$ & L4 & 0.03 & 8.2 & 7.56 & 113021 &  21.97 & 0.59 $\pm$ 0.045 & 0.16 $\pm$ 0.035 & 0.17 $\pm$ 0.058 & red\\
          &  & && & 120222 & 21.88 & --- & --- & 0.24 $\pm$ 0.055 & indeterminate \\\hline
         2007 VL$_{305}$$^{1,2}$ & L4 & 0.07 & 28.1 & 8.51 & 113021 &  22.60 & 0.60 $\pm$ 0.054 & 0.25 $\pm$ 0.038 & -0.15 $\pm$ 0.109 & red\\
          & & & & & 120222  & 22.60 & --- &---& 0.30 $\pm$ 0.047 & indeterminate\\\hline
         2010 TS$_{191}$$^{3}$ & L4& 0.05 & 6.6 & 8.07 & 113021 &  22.39 & 0.61 $\pm$ 0.029 & 0.30 $\pm$ 0.029 & 0.64 $\pm$ 0.078 & red\\ \hline
         2011 SO$_{277}$$^{3}$ & L4& 0.01 & 9.6& 7.76 & 113021  & 22.43 & 0.60 $\pm$ 0.067 &--- & ---- & indeterminate\\ 
         &&&& & 120222 & 22.53 & --- & 0.57 $\pm$ 0.050 & 0.82 $\pm$ 0.047 & ultra-red\\\hline
         2012 UD$_{185}$$^{5}$ & L4& 0.04& 28.3& 7.59& 113021  & 22.32 & 0.61 $\pm$ 0.033 & 0.37 $\pm$ 0.045 & 0.12 $\pm$ 0.081 & red \\ \hline
         2012 UV$_{177}$$^{5}$ & L4&0.07 & 20.8& 9.28 & 113021  & 23.76 & 0.71 $\pm$ 0.058 & 0.23 $\pm$ 0.051 & --- & red\\ \hline
         2013 RL$_{124}$$^{5}$ & L4& 0.03 & 10.1& 8.83& 113021  & 23.37 & 0.38 $\pm$ 0.075 & 0.54 $\pm$ 0.086 & 0.67 $\pm$ 0.128 & red\\\hline
         2013 TZ$_{187}$$^{5}$ & L4& 0.07& 13.1& 8.19 & 113021  & 23.27 & 0.90 $\pm$ 0.053 & 0.30 $\pm$ 0.057 &---& ultra-red\\\hline
         2013 VX$_{30}$$^{4,5}$ & L4& 0.09 & 31.2& 8.31 & 113021  & 22.60 & 1.01 $\pm$ 0.043 & 0.44 $\pm$ 0.043 & 0.86 $\pm$ 0.049 & ultra-red\\
          & && && 091122& 22.96 & 0.70 $\pm$ 0.104 & 0.47 $\pm$ 0.048 & 0.73 $\pm$ 0.045 & ultra-red\\\hline
         2014 RO$_{74}$$^{5}$& L4& 0.05 & 29.5& 8.39 & 120222  & 23.34 & 0.65 $\pm$ 0.052 & 0.42 $\pm$ 0.064 & 1.42 $\pm$ 0.069 & ultra-red\\\hline
         2014 SC$_{374}$$^{5}$ & L4 & 0.10 & 33.7 & 8.18& 113021  & 23.24 & 0.43 $\pm$ 0.066 & 0.12 $\pm$ 0.081 &---& blue\\\hline 
         2014 YB$_{92}$${5}$ & L4& 0.10 & 30.8& 8.62 & 091222  & 23.41 & 0.46 $\pm$ 0.187 & 0.07 $\pm$ 0.100 & 0.36 $\pm$ 0.090 & blue\\ 
         \hline
         2015 VU$_{207}$$^{5}$& L4 & 0.03 & 38.9& 7.28  & 080922  & 22.23 & 0.31 $\pm$ 0.034 & 0.24 $\pm$ 0.031 & 0.40 $\pm$ 0.024 & red\\
          & &&&& 091122 & 22.10 & 0.47 $\pm$ 0.052 & 0.09 $\pm$ 0.068 & 0.35 $\pm$ 0.028 & blue\\\hline
         2015 VV$_{165}$$^{5}$ & L4 & 0.09 & 16.8 & 9.02& 113021  & 23.32 & 0.87 $\pm$ 0.049 & 0.32 $\pm$ 0.055 & ---& ultra-red\\\hline
         2015 VW$_{165}$$^{5}$ & L4& 0.05 & 5.0 & 8.39& 113021 & 22.89 & 0.45 $\pm$ 0.032 & 0.36 $\pm$ 0.048 & --- & red\\
          & &&& & 120222 & 22.93 & --- &--- & 0.61 $\pm$ 0.060 & indeterminate\\
         \hline
    \end{tabular}}

    \label{tab:target_list}
\end{table}

\section{Photometric Reduction}\label{photreduc}


\subsection{Calibration}
To calibrate the photometry of our IMACS observations, we cross-matched the in-frame background stars against PS1 sources \citep{Magnier2013}.
We first converted the PS1 griz photometry to the SDSS system using the transformation equations in \citet{Tonry2012}, and then selected the sources with $g-r$ between 0.25 and 2.0, $r-i$ between 0.0 and 0.8 as the reference sources. By solving the equation below using the apparent magnitude of the reference sources, we determined the photometric zeropoint of each frame:
\begin{equation}
m_{sdss} = m_{ins} + 2.5\log_{10}(\tau_{exp}) + m_0,
\end{equation}
where $m_{sdss}$ is the apparent magnitude of a specific band of the cross-matched reference sources, $m_{ins}$ is the instrumental magnitude of that specific band measured from the IMACS image, $\tau_{exp}$ is the exposure time, and $m_0$ is the photometric zeropoint of that frame.

After we determined the zeropoints of each frame, we used every cross-matched star in every frame to evaluate the linear color conversions between the IMACS and SDSS photmetric system  by solving the following equation:
\begin{equation}
m_{M}  = m_{sdss} + a~(g-r)_{sdss} + b,
\end{equation}
where $m_{M}$ and $m_{sdss}$ are the IMACS and SDSS magnitude, respectively, and a, b are the coefficients of the linear conversion. The results are:
\begin{equation}
\label{eq:transformation}
\begin{split}
g_{M} = g_{sdss} -0.078(g-r)_{sdss} + 0.069 \\
r_{M} = r_{sdss} -0.024(g-r)_{sdss} + 0.024 \\
r_{M} = r_{sdss} -0.038(r-i)_{sdss} + 0.015  \\
i_{M} = i_{sdss} -0.188(r-i)_{sdss} + 0.134 \\
z_{M} = z_{sdss} -0.026(g-r)_{sdss} + 0.031
\end{split}
\end{equation}
With the photometric zeropoints and the color conversion equations, we are able to measure the griz colors of targets in SDSS photmetry system.

\subsection{PSF Modeling}
\begin{figure}
    \includegraphics[scale=.7]{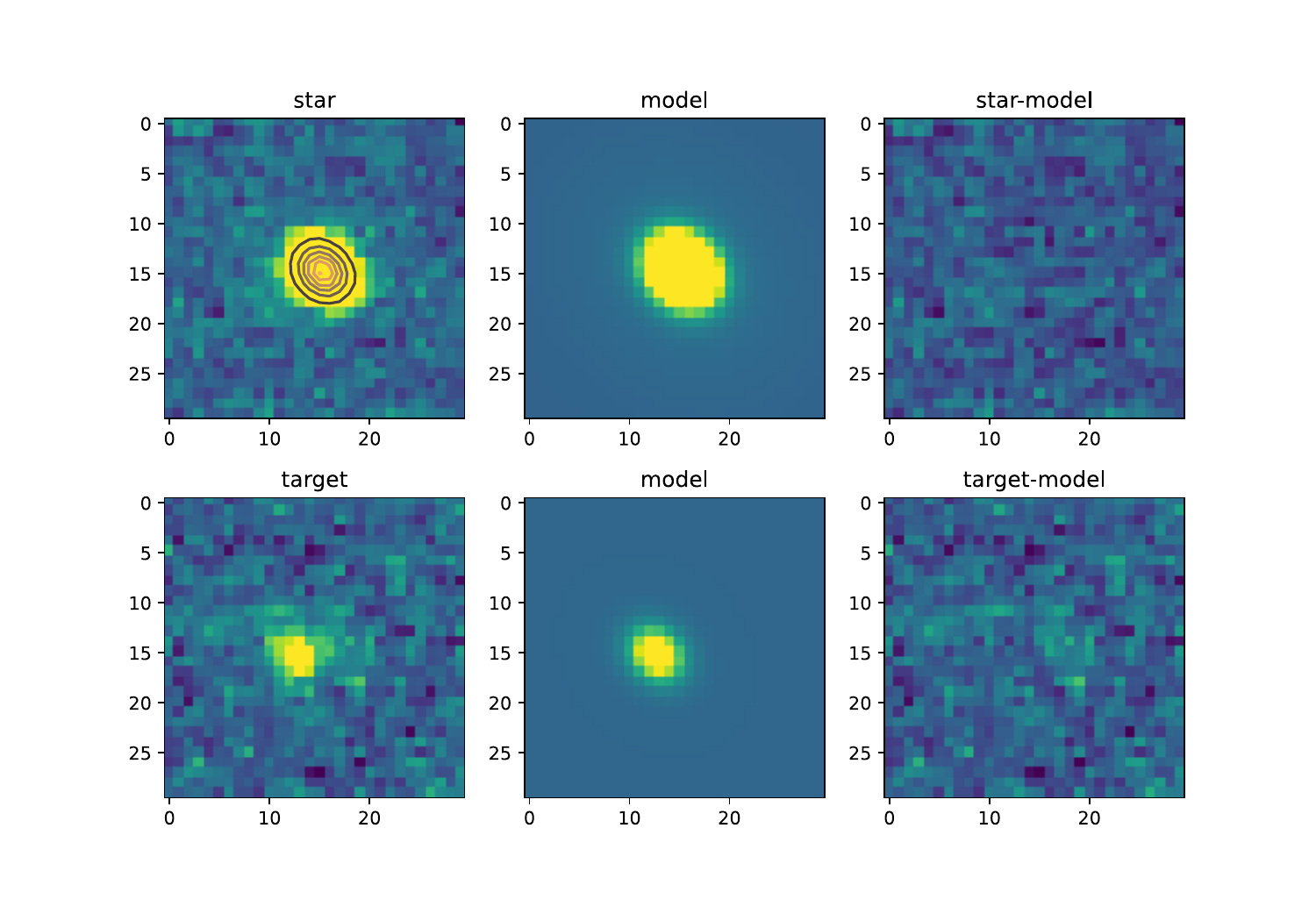}
    \caption{PSF modeling and subtraction. \textbf{top-left:} A star with the PSF model contour. \textbf{bottom-left:} The image of NT. \textbf{middle:} The model of the star (top) and the NT (bottom). \textbf{right:} the images after subtraction of the model.}
    \label{fig:psf}
\end{figure}

To accurately measure the flux and apparent magnitude of NTs, we select stars around the target NT to model the local PSF. Several popular analytical functions are considered for modeling the PSF, such as Moffat \citep{Moffat} and the sum of 2D Gaussians \citep{Bendinelli90}. Both functions can adequately model the ``wing'' of PSF. However, considering our PSF can be asymmetric (not round, see Figure~\ref{fig:psf}), we model the PSF by using the superposition of n asymmetric 2D Gaussians. The flux of the PSF at any point in the $(x',y')$ orthogonal coordinate system is:

\begin{equation}
PSF(x',y') = b(x', y') + 
\sum_{i=1}^{n}
\mathrm{A_i}~\Big(\texttt{exp}\big[-(\frac{x'^2}{2\sigma^{2}_{x'i}}+\frac{y'^2}{2\sigma^{2}_{y'i}})\big]\big),
\end{equation}
where $b(x', y')$ is the background flux at that point, n is a small number, A$_i$ is the amplitude of individual Gaussian, $\sigma_{x'i}$ and $\sigma_{y'i}$ are the widths on $x'$ and $y'$ axes of individual Gaussian, respectively. This equation can be rotated to the image reference frame $(x, y)$ with a position angle $\theta$ and translating the centroid to $(x_0,y_0)$ such that

\[\begin{pmatrix}x'\\y'\end{pmatrix}=   \begin{bmatrix}
     \cos\theta & -\sin\theta\\
     \sin\theta & \cos\theta
   \end{bmatrix}
   \begin{pmatrix}x-x_0 \\y-y_0 \end{pmatrix}. \]
Therefore, the Gaussian functions share the same center, position angle, and ellipticity but have unequal contribution and different width. To proper chose `n', the number of Gaussians we should use, we calculate Bayesian information criterion (BIC) for each n we use. The BIC is defined as:
\begin{equation}
\mathrm{BIC} = -2\: \mathrm{ln}(\hat{\mathcal{L}}) + k\: \mathrm{ln(m)},
\end{equation}
where $\hat{\mathcal{L}}$ is the maximum likelihood of the model, $k$ is the number of parameters estimated by the model, and m is number of data points we use to fit the model. The models with lower BIC values are generally preferred, and it penalizes the model with larger $k$ automatically. Since the multiple Gaussian PSF model can be linearized by taking logarithm and assuming that the errors are normally-distributed, the $\hat{\mathcal{L}}$ is equivalent to the least squares estimation. Thus, BIC can be written as a function of error variance $\hat{\sigma_e^2}$:

\begin{equation}
\mathrm{BIC} = m\: \mathrm{ln}(\hat{\sigma_e^2}) + k\: \mathrm{ln(m)},
\end{equation}
In other words, the model with lower residual and fewer parameters is preferred.  
We find that the model with n = 1, a single 2D Gaussian, always has highest BIC. On the other hand, the models with n = 2 and n = 3 generally have similar BICs, therefore we conclude that using any model with n $>$ 3 is redundant. 

Finally, we use the PSF model with n = 2 or 3, depending on which one has lower BIC. Once all of the parameters are measured via modeling the stars, the target NT can be modeled by refitting the center and amplitude of the PSF. The flux is the sum of the final model. Figure~\ref{fig:psf} demonstrates that both the star and the NT can be properly subtracted by the PSF model. 

\subsection{Rotation Curve Correction}\label{lightcurve}
\begin{figure}
    \centering
    \includegraphics{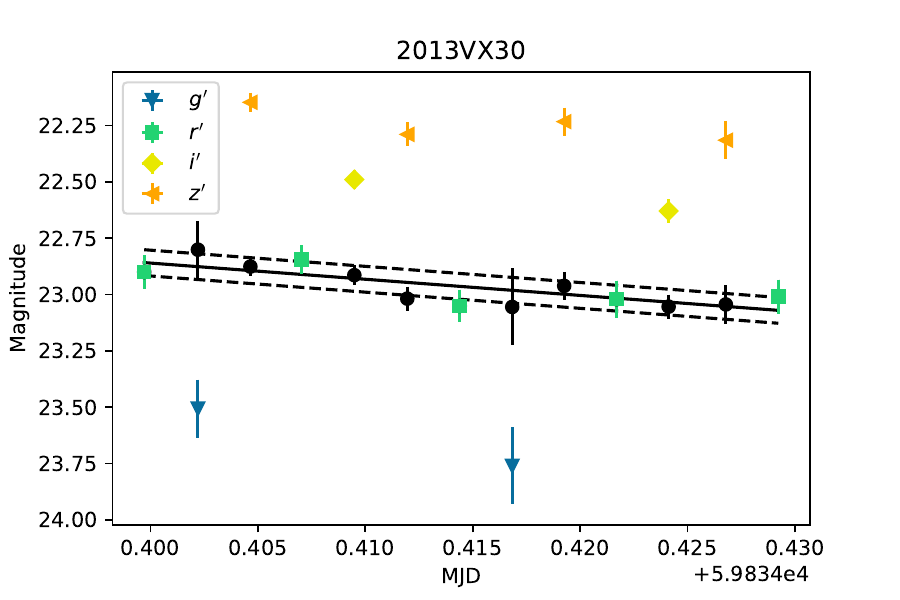}
    \caption{This figure shows our least-squares approach to fitting $r^{\prime}$-band lightcurves and colors for an example NT target, 2013 VX30. Each observation is shown as a colored point (blue downward triangle for $g^{\prime}$, green square for $r^{\prime}$, yellow diamond for $i^{\prime}$, and orange sideways triangle for $z^{\prime}$}. We then used a constant, but free parameter, color term to convert each observation to an $r^{\prime}$-band observations; these point are shown as black circles. The solid line is our least-squares fit to the $r^{\prime}$-band (black) points. The dotted lines show our 1$\sigma$ deviation from this fit. 
    \label{fig:lightcurve}
\end{figure}

The observed magnitudes, and the resulting colors we are trying to measure, are subject to rotational variations on the surface of these objects. To approximately account for this, we use a model that exhibits a linear variation in source brightness (its $r^{\prime}$-band magnitudes) and constant $g^{\prime}-r^{\prime}$, $r^{\prime}-i^{\prime}$, $r^{\prime}-z^{\prime}$ colors (to convert each measurement to and $r^{\prime}$-band magnitude). This model was then fit using a least-squares approach (see Fig. \ref{fig:lightcurve}). The resulting colors have been converted to SDSS magnitudes ($griz$; see Eq. \ref{eq:transformation}), which are reported in Table \ref{tab:target_list}. 

\subsection{Reddening Line}\label{redline}
Taken from \cite{2002A&A...389..641H}, the reddening, or the spectral index, can be expressed as the percent of reddening per 100 nm:
\begin{equation}
    S(\lambda_1, \lambda_2) = 100 * \frac{R(\lambda_2) - R(\lambda_1)}{(\lambda_2 - \lambda_1) / 100}
\end{equation}
where $R(\lambda)$ is taken from \cite{1986ApJ...310..937J}:
\begin{equation}
    R(\lambda) = 10 ^ {-0.4 (m(\lambda) - m_\odot(\lambda))}
\end{equation}
such that $m(\lambda)$ and $m_\odot(\lambda)$ are the magnitude of the object and the Sun, respectively, at a particular wavelength, $\lambda$. Setting the reddening line to pass through the color of the Sun (i.e. for $S(\lambda_1, \lambda_2)$ = 0, $m(\lambda_1) - m(\lambda_2) = m_\odot(\lambda_1) - m_\odot(\lambda_2)$), we can derive the following equation, assuming $m(\lambda_1) = m_\odot(\lambda_1)$:
\begin{equation}
    m(\lambda_2) = -2.5 log[1 - 10^{-4} S(\lambda_1, \lambda_2) (\lambda_1 - \lambda_2)] + m_\odot(\lambda_2)
\end{equation}
Assuming $S(\lambda_1, \lambda_2)$ varies from -10\% to 80\%, we can plot the reddening line for $g-r$ vs $r-i$ and $g-r$ vs $r-z$ in Fig. \ref{fig:gr_ri} and Fig. \ref{fig:gr_rz} respectively. Note that our targets generally fall along the reddening line, as has been observed for small bodies in the outer Solar System previously \citep{2002A&A...389..641H}. Objects that fall above/below the reddening line must exhibit emission/absorption lines at those particular wavelengths causing them to deviate from a flat spectral index.

\section{Results}
\subsection{Color-Color Results}

\begin{figure}
    \centering
    \includegraphics{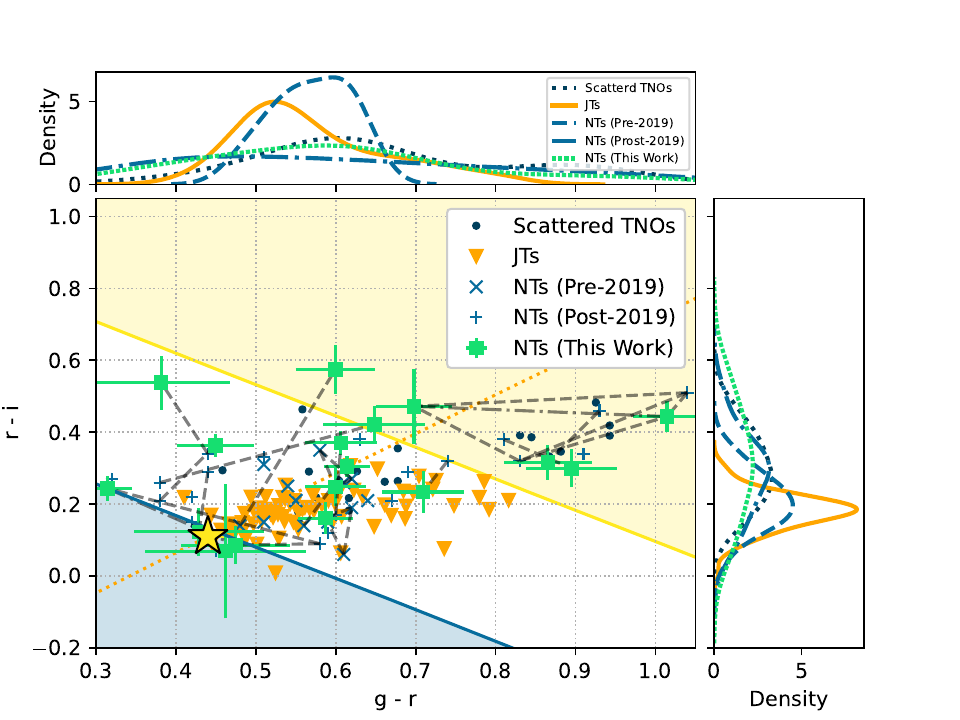}
    \caption{Measured $g-r$ vs $r-i$ of the NT population. Blue points are colors of scattered TNOs and orange triangles are JTs, both taken from the literature \citep{2012A&A...546A.115H}. Light blue x's are previously observed colors of NTs which the "Trojan Color Conundrum" was based on \citep{2006Sci...313..511S, 2012AJ....144..169S, 2013AJ....145...96P, 2018AJ....155...56J}, while the blue plus signs are more recently observed NT colors which bring this conundrum into question \citep{2019Icar..321..426L, Bolin2023}. Targets observed in this paper are shown as green squares. Solar color and the reddening line (see \ref{redline}) are depicted as an yellow star and orange dotted line respectively. Objects that have multiple observations in this paper are connected by a dot-dashed line. NTs that have been previously observed in the literature are connected by a dashed line. The yellow line marks values where the PCA yields values equal to our cutoff of 1.75 (see Fig. \ref{fig:pca_results} and Sec. 4.2). Objects in the yellow region are above this cutoff and considered ultra-red in this paper. The blue line marks values where the PCA yields values equal to our cutoff of -1.25 (see Fig. \ref{fig:pca_results} and Sec. 4.2). Objects in the blue region are blue this cutoff and considered blue in this paper. The top and right inset plots show the kernel density estimation (KDE) of the g-r and r-i distributions respectively of the included sub-populations.}
    \label{fig:gr_ri}
\end{figure}

In Fig. \ref{fig:gr_ri}, we show the $g-r$ and $r-i$ colors measured for our NT targets. Similar to the scattered TNOs, our targets exhibit wide range in this color space; while most targets fall within the ``red" zone (principal component \textless 1.75; see Sec. 4.2), there are three firm and one potential NTs in the ``ultra-red" zone (principal component \textgreater 1.75). Of these objects, we identified two new ultra-red NTs, 2013 TZ$_{187}$ and 2015 VV$_{165}$, which were also independently found and reported in \citet{Bolin2023}. The potential ``ultra-red" NT, 2011 SO$_{277}$ has varying results from different observations (\citet{2018AJ....155...56J}, \citet{2019Icar..321..426L}, and this work); see more discussion of this object in Sec. 4.4. 

With extra ultra-red colored NTs, the red to ultra-red ratio for our sample is 3.75:1, or 7.75:1  for the entire known population. This ratio is much more consistent with the dynamically excited KBO ratio of between 4-11 : 1 \citep{2019ApJS..243...12S}. However, comparing these ratios is not sufficient to determine if the NT and KBO population come from the same source distribution (see Sec. 4.2). We also show the kernal density estimations (KDEs) of g-r and r-i color in Fig. \ref{fig:gr_ri}. Unlike the results from previous works, which claimed that the NTs and JTs have very similar color distributions, our new results show that the KDEs of NTs are closer the the KDEs of scattered TNOs. Further analysis is presented in Sec. 4.2.

\begin{figure}
    \centering
    \includegraphics{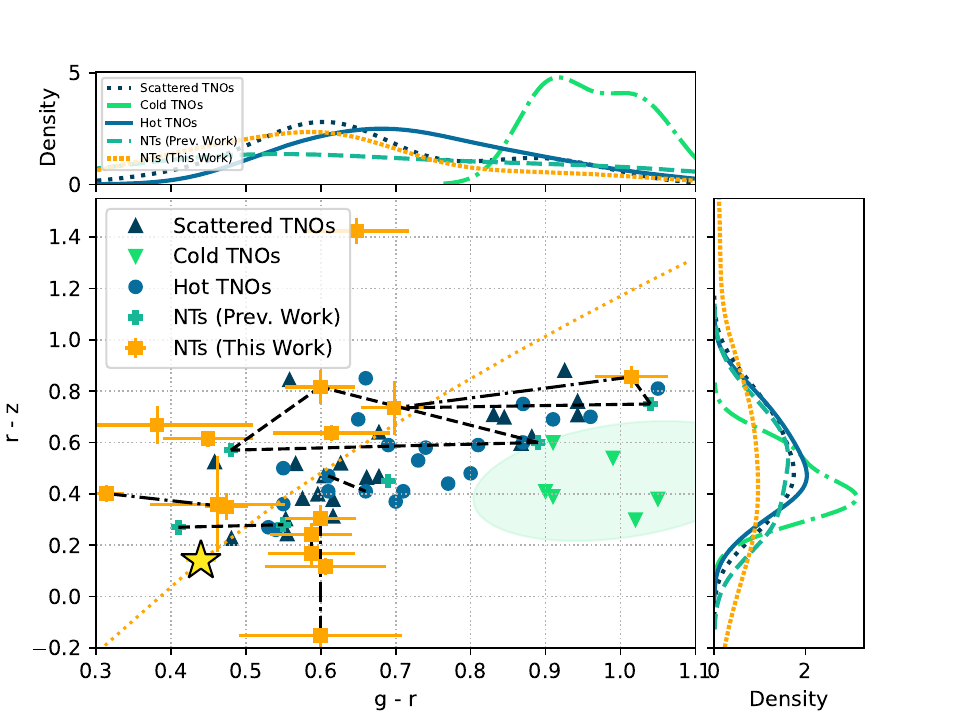}
    \caption{Measured $g-r$ vs $r-z$ of the NT population. Navy upward triangles, green downward triangles, and blue circles are measurements taken from the literature of TNOs (scattered, cold, and hot respectively) \citep{2019ApJS..243...12S}. Teal plus signs are colors of NTs taken from the literature \citep{2019Icar..321..426L}. Targets observed in this paper are shown as orange squares. Solar color and the reddening line (see \ref{redline}) are depicted as an yellow star and orange dotted line respectively. Objects that have observations taken in this paper and from the literature are connected with a dashed line. Objects that have multiple observations in this paper are connected by a dot-dashed line. The green ellipse demarcates the region of color-color space occupied only by cold TNOs. The top and right inset plots show the kernel density estimation (KDE) of the g-r and r-z distributions respectively of the included sub-populations.}
    \label{fig:gr_rz}
\end{figure}

In Fig. \ref{fig:gr_rz}, we show the $g-r$ and $r-z$ colors measured for our NT targets. All of our targets are consistent with the scattered/hot TNO populations. This result is expected as NTs are thought to have originated from scattered/unstable TNOs. The physical cause of this $z$-band colorization of the cold TNO population is not currently clear, but must be due to some absorption feature around 900 nm based on the displacement from reddening line. Spectroscopic information, such as will be taken with JWST \citep{2021jwst.prop.2550M}, will shed further light on chemical links between these populations. 

\subsection{Comparison to Previous Observations}

\begin{figure}
    \centering
    \includegraphics{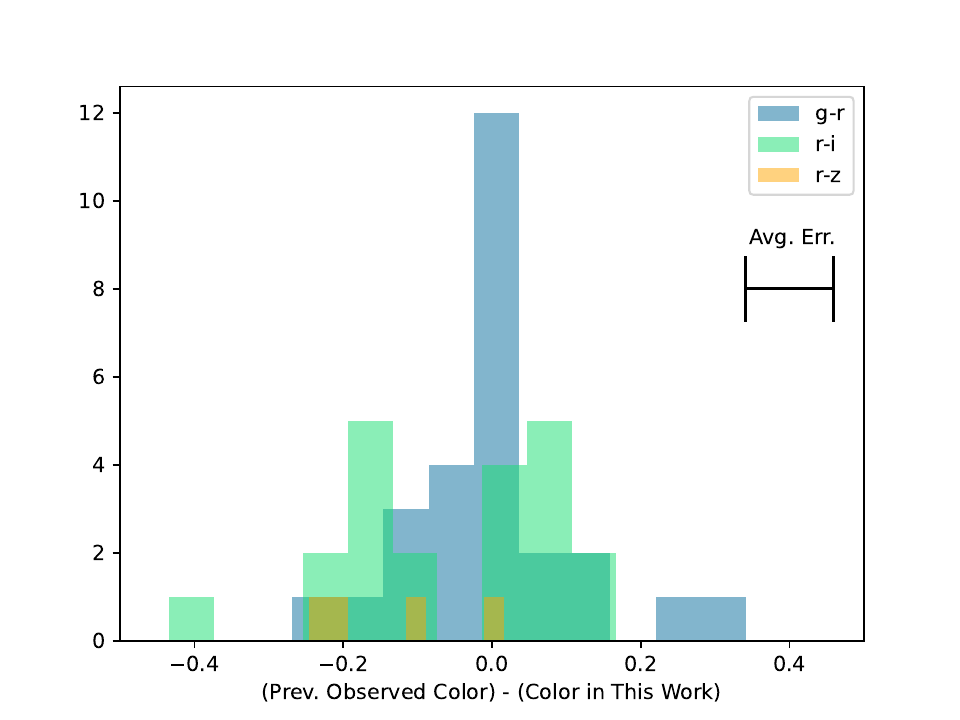}
    \caption{The differences in observed color between NTs in this paper and the literature as compared to the average error on our observations. The differences in g-r, r-i, and r-z observations are shown as blue, orange, and green histograms respectively. The average g-r, r-i, and r-z errors are shown as blue dotted, green dot-dashed, and orange dashed lines respectively}
    \label{fig:errors}
\end{figure}

All of the targets in our sample have previous observations (though not all from the same survey). Therefore, we compare the difference between our measurements and those from the literature to our computed errors, shown in Fig. \ref{fig:errors}, to determine if there is any systematic offset in our observations. We find that the observed differences in g-r are mostly within our observational errors, meaning our observations are roughly consistent with previous literature. However, previous observations are split between being slightly systematically larger in r-i and systematically lower than our measurements. Further investigation indicated that the larger group has smaller offset in the order of 0.05, and the lower group has larger offset about -0.15. We also find an instrument dependency on the groups; the smaller offset samples were mostly measured with Gemini and Dark Energy Survey, which both have proper photometry transformation equations to SDSS system. On the other hand, the larger offset samples were mostly measured by using the R and I filters or without proper photometry transformation equations. Therefore it is likely that the different photometry systems mostly contribute such systematic offsets. 
In every case, this did not change the result much on the following Principal Component Analysis (PCA) analysis, since the g-r axis is the dominant element on our Principal Component. 

\subsection{Comparison to Other Populations}
\begin{figure}
    \centering
    \includegraphics{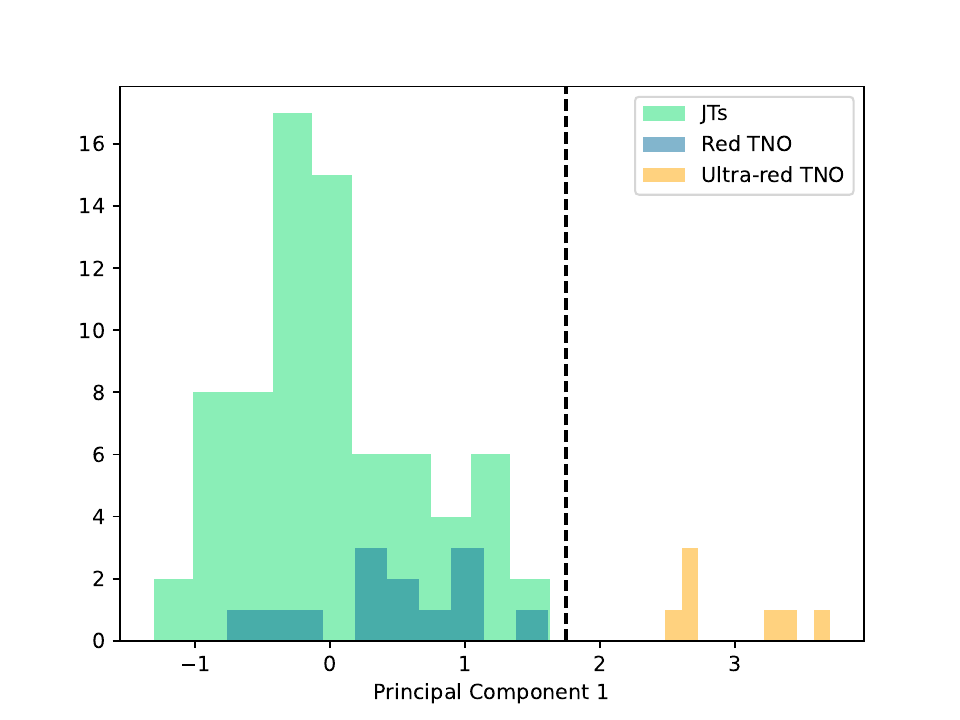}
    \caption{The results of running a Principal Component Analysis (PCA) with the g-r and r-i colors of certain Solar System populations. The green histogram corresponds to the JTs (taken from \cite{2012A&A...546A.115H}). The blue and orange histograms correspond to the red and ultra-red subpopulations of the scattered TNOs, taken from \citet{2012A&A...546A.115H}; the classification of red vs ultra-red was determined by using a clustering algorithm (DBSCAN; \citet{scikit-learn}) which separated the TNOs into two sub-populations.}
    \label{fig:pca_results}
\end{figure}

\begin{figure}
    \centering
    \includegraphics{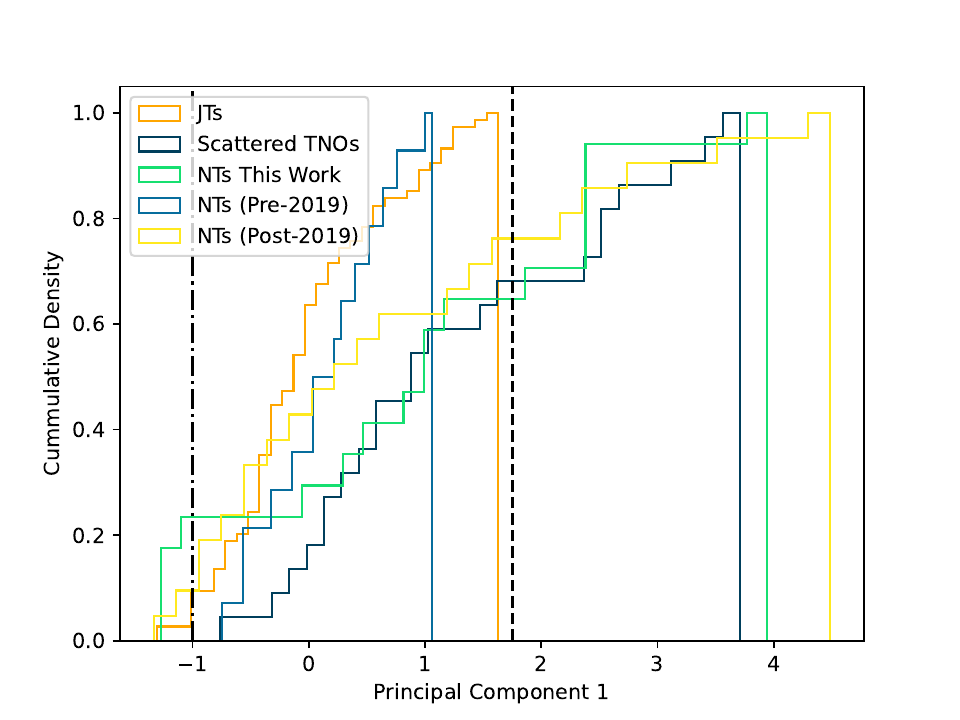}
    \caption{Cumulative distributions of the Principal Component (see Sec. 4.2) values of populations in the Solar System. The cut-off between red and ultra-red as defined by this PCA is shown as a black dashed line (see Fig. \ref{fig:pca_results}). The cut-off between red and blue objects is similarly shown as a dot-dashed line. The JT and scattered TNO results are shown as orange and navy histograms respectively. The NTs observations from previous literature are shown as a blue histogram. The NT observations from this work are shown as a green histogram. }
    \label{fig:cumulative_dists}
\end{figure}

The ultimate goal of this work is to determine how similar the NT colors are to other populations in the Solar System. A simple statistical test to measure the likelihood that two distributions are drawn from the same underlying distribution is the Kolmogorov-Smirnov (KS) test \citep{10.2307/2237048}. Although the KS test can be generalized to more 
than a single dimension, the interpretation becomes complicated. For simplicity, 
we reduce the dimension of our data and use the traditional statistical test. 
Specifically, we performed a Principal Component Analysis (PCA) of our data, using the scikit-learn python package \citep{scikit-learn}. Fig. \ref{fig:pca_results} demonstrates that the PCA is able to successfully reduce the g-r vs. r-i color-color plot to a 1-D parameter that still distinguishes between the red and ultra-red populations of TNOs and the whole JT population (which is comprised of only red objects). The principal component value (PC1) which separates these populations is 1.75 (shown as a dotted line in Fig. \ref{fig:pca_results}). We use this definition to classify our NT targets as red or ultra-red; the corresponding region in g-r vs r-i space is shown in Fig. \ref{fig:gr_ri} as a yellow shaded region. We then applied this PCA model to other populations in the Solar System, including JTs and previous observations of NTs, the results of which are shown in Fig. \ref{fig:cumulative_dists}. By eye, the JT population is clearly unique in that it is nearly devoid of any ultra-red members (i.e. targets with a PC1 \textgreater 1.75). Also of note, about 25\% of the NT targets presented in this paper occupy a unique region of PC1 $\sim -1$. This region corresponds to blue objects that are not frequently present in the outer Solar System populations (see Sec. 4.4 for a more in-depth discussion of these objects). 

\begin{table}
    \centering
    \begin{tabular}{c|c|c|c|c|c}
         KS Test P-value & NTs (This Work) & NTs (Pre-2019) & NTs (Post-2019) &TNOs & JTs \\ \hline
         NTs (This Work) & 1 & 0.020 & 0.61 & 0.56 & 0.003\\ \hline 
         NTs (Pre-2019) & 0.020 & 1 & 0.15 & 0.03 & 0.27 \\ \hline
         NTs (Post-2019) & 0.61 & 0.15 & 1 & 0.14 & 0.05 \\ \hline
         TNOs & 0.56 & 0.03 & 0.14 & 1 & 0.0002\\ \hline
         JTs & 0.003 & 0.27 & 0.05 & 0.0002 & 1\\ \hline
    \end{tabular}
    \caption{The resulting p-values of the KS Test on each combination of sub-populations considered in this work.}
    \label{tab:ks_test}
\end{table}

We then ran a KS test for each combination of these Solar System populations to determine the likelihood that they came from the same underlying distribution; the results of these tests are recorded in Table 2. We conclude that the compared populations are from different distributions if they have a p-value of $\le$ 0.05, corresponding to a 95\% confidence level to reject the null hypothesis. Therefore, we find that the population observed in this work is not consistent with being drawn from the same distribution as the JTs, but is instead more consistent with the TNO population. This result is the opposite of what was found pre-2019, where the NTs were more consistent with the JT population. The results from post-2019 data also show that the NT population is more consistent with the TNO population, but this work shores up this result significantly. Further observations of members of the NT population in particular could also increase the statistical significance of this result. However, we feel confident in claiming that our results show NTs and TNOs are consistent with coming from the same underlying distribution based on their optical colors with the greatest confidence to date.

\subsection{Color-Absolute Magnitude Relations}

\begin{figure}
    \centering
    \includegraphics[width=\textwidth,height=\textheight,keepaspectratio]{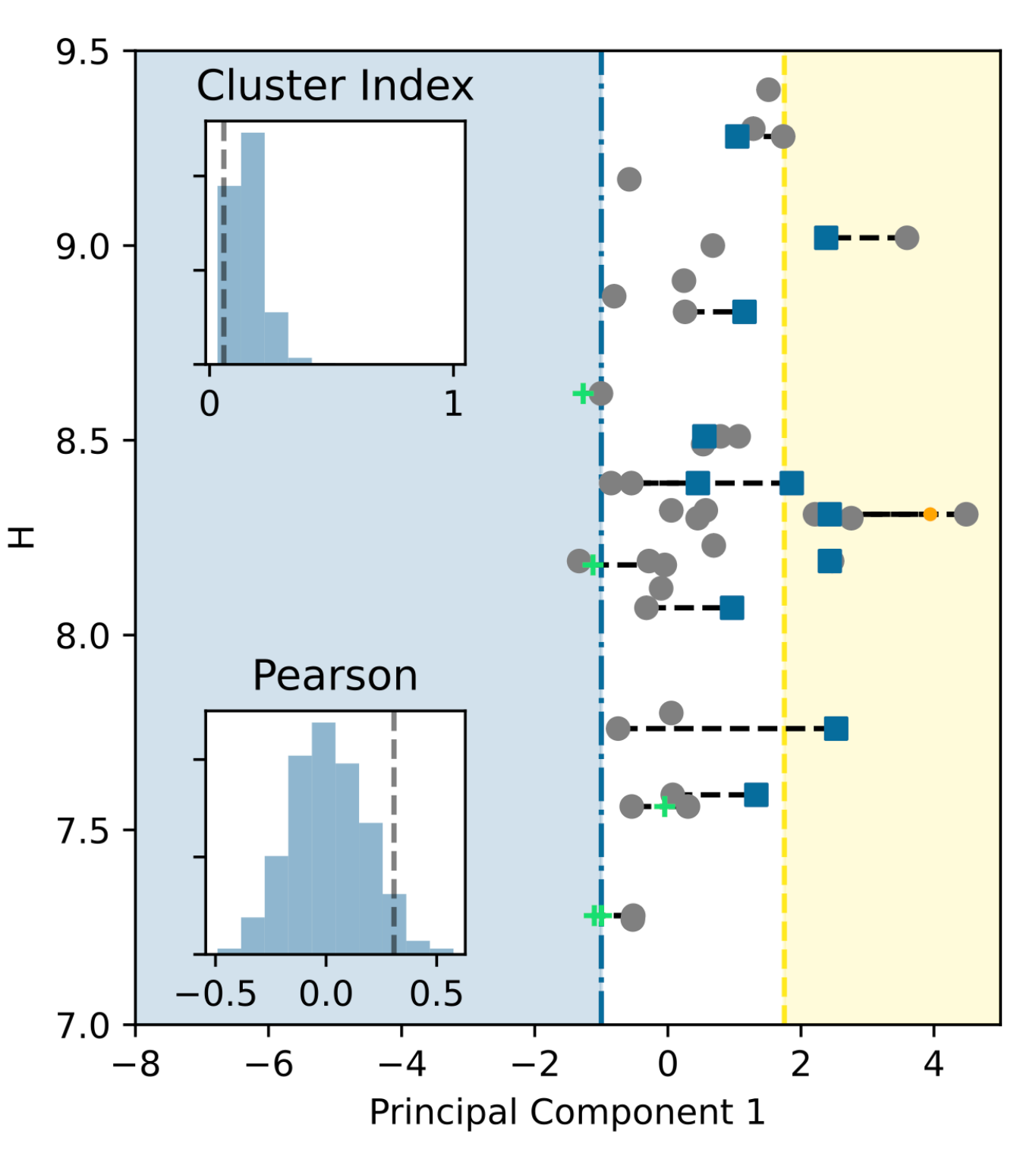}
    \caption{NT colors as a function of absolute magnitude. Grey points are taken from the literature \citep{2006Sci...313..511S, 2012AJ....144..169S, 2013AJ....145...96P, 2018AJ....155...56J, 2019Icar..321..426L, 2019ApJS..243...12S}. Colored squares were measured in this paper. Duplicate observations of the same object are connected by dashed lines. The inset plots contain histograms of the Cluster Indices and Pearson Correlation Coefficients of a random distribution colors and  absolute magnitude (see Sec. 4.1). Each grey dashed line in the inset plots shows the corresponding value calculated for the observed distribution.}
    \label{fig:orbital_elements}
\end{figure}

In Fig. \ref{fig:orbital_elements}, we plot the Principal Component for our targets as a function of 
absolute magnitude (H). 
We look for any significant clustering or correlations in these plots which would indicate that the color classification of NTs is dependent on their size.

To search for clustering in our datasets, we run a Mean Shift Clustering algorithm \citep{scikit-learn}, which does not need a number of clusters as an input parameter (just a bandwith which can be initialized with the \texttt{estimate\_bandwith} function). To test the significance of clustering we calculate the Cluster Index. The Cluster Index from the SigClust evaluation tool is defined as \citep{2013RMxAA..49..137F}:
\begin{equation} 
    CI = 
     \frac{\sum_{k=1}^{N} \sum_{i \in C_k}\parallel \pmb{x_i} - \pmb{\Bar{x}} ^ {(k)} \parallel ^ 2} 
     {\sum_{i =1 }^{n}\parallel \pmb{x_i} - \pmb{\Bar{x}} \parallel ^ 2}
\end{equation}
where $ \pmb{\Bar{x}} ^ {(k)}$ represents the mean of the kth cluster for k = 1, 2, ... N for N clusters and  $\pmb{\Bar{x}}$ represents the overall mean. The CI provides a p-value for the significance of the cluster between these two clusters. To test if our data was correlated, we used the Pearson Correlation Coefficient \citep{ref1} which is defined as:
\begin{equation}
    r = \frac{\sum (x_i - \Bar{x}) (y_i - \Bar{y})}{\sqrt{\sum (x_i - \Bar{x})^2 \sum (y_i - \Bar{y})^2}}
\end{equation}
where $x_i$ and $y_i$ are the data points and $\Bar{x}$ and $\Bar{y}$ are the respective means. We calculated each of these values for all of the plots shown in Fig. \ref{fig:orbital_elements}. To determine whether or not these values could be obtained from random noise we generated 1000 sets of points with the same number of objects as our observation within the same region of Principal Component vs H space and ran the same analysis on those sets. These results are shown in the inset histograms in Fig. \ref{fig:orbital_elements}.

We found that the cluster is consistent with random noise and should not be considered significant. This result also suggest that the color of NTs are distributed continuously from blue to ultra-red rather than bimodally.  
The positive correlation with size is intriguing and may point to primordial differences in objects of different sizes in the outer Solar System. However, H is not a direct correlation to size as the object's albedo must be taken into account. Such observations do not currently exist for the NT population and will be necessary to establish a color-size correlation. Indeed, photometric observations of the rest of the NT population are necessary to confirm this slight correlation.

\subsection{Unique Targets}

While most of our targets are consistent with previous color measurements, one object, 2011~SO$_{277}$ is classified here as ultra-red while its previous observations place it firmly with in the red zone. Based on our other observations, we consider our results to be roughly consistent with previous literature (see Fig. \ref{fig:errors}), so this result is indeed unexpected. One explanation as to why this object has such different colors in independent observations is that its surface is not homogeneous. To test this hypothesis, a more in depth study of the rotational properties of the surface of this object is necessary, which will be upcoming in our next work on the lightcurves of NTs.

Three of our targets, 2014~SC$_{375}$, 2014~YB$_{92}$, and 2015~VU$_{207}$, are much bluer, nearly solar in color, as compared to the other NTs or KBOs. \citet{Bolin2023} also reported 2014~YB$_{92}$ and 2015~VU$_{207}$ have blue, near solar color.
In fact, these objects are as blue as the blue B/C-type asteroids, such as 3200 Phaetheon \citep{Tabeshian_2019, LISSE2022114995}. A similarly blue TNO has been observed, which appears to be covered ferric oxides and phyllosilicates \citep{2018ApJ...855L..26S}. This TNO has a highly eccentric and inclined orbit, suggesting it may have a common origin with C-type asteroids and has since been implanted into trans-Neptunian space. It is possible that these NTs originated elsewhere in the Solar System, but their current orbits are stable for \textgreater\ Gyrs (see Sec. \ref{survey}), implying that they were captured just after Neptune's migration. However, based on these results the blue ratio for NTs is currently much higher than that of the TNO population. This result may suggest that inner Solar System material may be more efficiently transferred to NT orbits which have a smaller perihelion than the Kuiper Belt. Future spectral observations would be necessary to reveal any compositional differences this target may have as compared to the rest of the NT population.

\section{Why were the ultra-red NTs rare before 2019?}

Prior to 2019, the ultra-red NTs were very rare; none of the 13 NT samples in \citet{2018AJ....155...56J} are ultra-red NTs, which led to the claim of a ``Trojan Color Conundrum''. Here we propose two possibilities to explain this inconsistency:
\begin{enumerate}
    \item \textbf{Small number statistics:} Small number statistics could generate such a surprising result.
    If we assume a 7.75:1 apparent red to ultra-red ratio of NTs, the chance to randomly select 13 objects without picking up any ultra-red one is about 18\%, which is very likely. If we use a 3.75:1 apparent red to ultra-red ratio, the chance is now 0.5\%. While it is not impossible, we may also consider alternative explanations.  
    \item \textbf{Selection effect:} Since bigger objects are easier to detect and obtain color measurements for, the 13 objects in \citet{2018AJ....155...56J} trend to be large; 10 of 13 have H $\leqslant 8$. Moreover, many NTs have been discovered by deeper \citep{2021Icar..36114391L} or wider \citep{2022ApJS..258...41B, 2020ApJS..247...32B, 2019Icar..321..426L} surveys since 2018, which included many high-inclination objects. Thus the \citet{2018AJ....155...56J} sample appears to be biased toward bigger sized and lower inclination objects. In fact, 8 of 13 NTs in the \citet{2018AJ....155...56J} sample have orbital inclination $< 10^{\circ}$; 9 of the 31 currently known NTs have inclination $< 10^{\circ}$, meaning that 8 of the 9 total low-inclination NTs were included in the \citet{2018AJ....155...56J} sample. Such objects has very similar red color (see Figure~\ref{fig:orbital_elements}). Therefore, the possible color-orbit-size correlation in NT population could be at least partially explain why the ``Trojan Color Conundrum'' was observed, especially when there were some selection biases in that sample.
\end{enumerate}

\section{Conclusions}
In this paper, we measure the griz colors for 15 of the 24 known NTs. We used the IMACS f/4 instrument on the 6.5m Baade telescope with Sloan g'r'i'z' filters to conduct our photometric survey. We confirm that 2013~VX$_{30}$ is ultra-red in color, and identify three NTs as ultra-red. This result brings the red to ultra-red ratio of NTs to 7.75:1, much more consistent with the corresponding TNO ratio and resolving the ``Trojan Color Conundrum". Moreover, the color distribution of NTs is now indistinguishable from the scattered population of TNOs and different from the Jovian Trojans.
We also find three targets which have solar color, the origin of which is unclear; the most likely explanation is that these objects originated from the inner solar system. For the entire NT population, we find that color of NTs may correlated to their absolute magnitude, and the objects with larger H trend to have redder color. The explanation behind this correlation remains an open question that is difficult to address with current data. 
More discoveries of NTs (especially around L5) are clearly needed. The L5 point has historically been difficult to study due to its overlap with the galactic plane, but the NT L5 region is moving away from this high stellar density region, making now the perfect time to start studying this population. The true degree of asymmetry between the L4 and L5 clouds will be an important to distinguishing different formation scenarios for the NT population. Moreover, our ongoing work to measure the rotational period and specific composition of these small bodies directly will be vital to understanding the true origin of the NT population. 

\begin{acknowledgments}
This paper includes data gathered with the 6.5 meter Magellan Telescopes located at Las Campanas Observatory, Chile.

This material is based upon work supported by the National Aeronautics and Space Administration under grant No.\ NNX17AF21G issued through the SSO Planetary Astronomy Program and by the National Science Foundation under grant No.\ AST-2009096. This research was supported in part through computational resources and services provided by Advanced Research Computing at the University of Michigan, Ann Arbor.
\end{acknowledgments}


\bibliography{sample631}{}

\begin{thebibliography}{}
\expandafter\ifx\csname natexlab\endcsname\relax\def\natexlab#1{#1}\fi
\providecommand{\url}[1]{\href{#1}{#1}}
\providecommand{\dodoi}[1]{doi:~\href{http://doi.org/#1}{\nolinkurl{#1}}}
\providecommand{\doeprint}[1]{\href{http://ascl.net/#1}{\nolinkurl{http://ascl.net/#1}}}
\providecommand{\doarXiv}[1]{\href{https://arxiv.org/abs/#1}{\nolinkurl{https://arxiv.org/abs/#1}}}

\bibitem[{{Almeida} {et~al.}(2009){Almeida}, {Peixinho}, \&
  {Correia}}]{2009A&A...508.1021A}
{Almeida}, A.~J.~C., {Peixinho}, N., \& {Correia}, A.~C.~M. 2009, \aap, 508,
  1021, \dodoi{10.1051/0004-6361/200911943}

\bibitem[{{Bendinelli} {et~al.}(1990){Bendinelli}, {Parmeggiani}, {Zavatti}, \&
  {Djorgovski}}]{Bendinelli90}
{Bendinelli}, O., {Parmeggiani}, G., {Zavatti}, F., \& {Djorgovski}, S. 1990,
  \aj, 99, 774, \dodoi{10.1086/115373}

\bibitem[{{Bernardinelli} {et~al.}(2020){Bernardinelli}, {Bernstein}, {Sako},
  {Liu}, {Saunders}, {Khain}, {Lin}, {Gerdes}, {Brout}, {Adams}, {Belyakov},
  {Somasundaram}, {Sharma}, {Locke}, {Franson}, {Becker}, {Napier},
  {Markwardt}, {Annis}, {Abbott}, {Avila}, {Brooks}, {Burke}, {Carnero Rosell},
  {Carrasco Kind}, {Castander}, {da Costa}, {De Vicente}, {Desai}, {Diehl},
  {Doel}, {Everett}, {Flaugher}, {Garc{\'\i}a-Bellido}, {Gruen}, {Gruendl},
  {Gschwend}, {Gutierrez}, {Hollowood}, {James}, {Johnson}, {Johnson},
  {Krause}, {Kuropatkin}, {Maia}, {March}, {Miquel}, {Paz-Chinch{\'o}n},
  {Plazas}, {Romer}, {Rykoff}, {S{\'a}nchez}, {Sanchez}, {Scarpine}, {Serrano},
  {Sevilla-Noarbe}, {Smith}, {Sobreira}, {Suchyta}, {Swanson}, {Tarle},
  {Walker}, {Wester}, {Zhang}, \& {DES Collaboration}}]{2020ApJS..247...32B}
{Bernardinelli}, P.~H., {Bernstein}, G.~M., {Sako}, M., {et~al.} 2020, \apjs,
  247, 32, \dodoi{10.3847/1538-4365/ab6bd8}

\bibitem[{{Bernardinelli} {et~al.}(2022){Bernardinelli}, {Bernstein}, {Sako},
  {Yanny}, {Aguena}, {Allam}, {Andrade-Oliveira}, {Bertin}, {Brooks},
  {Buckley-Geer}, {Burke}, {Rosell}, {Carrasco Kind}, {Carretero}, {Conselice},
  {Costanzi}, {da Costa}, {De Vicente}, {Desai}, {Diehl}, {Dietrich}, {Doel},
  {Eckert}, {Everett}, {Ferrero}, {Flaugher}, {Fosalba}, {Frieman},
  {Garc{\'\i}a-Bellido}, {Gerdes}, {Gruen}, {Gruendl}, {Gschwend}, {Hinton},
  {Hollowood}, {Honscheid}, {James}, {Kent}, {Kuehn}, {Kuropatkin}, {Lahav},
  {Maia}, {March}, {Menanteau}, {Miquel}, {Morgan}, {Myles}, {Ogando},
  {Palmese}, {Paz-Chinch{\'o}n}, {Pieres}, {Malag{\'o}n}, {Romer}, {Roodman},
  {Sanchez}, {Scarpine}, {Schubnell}, {Serrano}, {Sevilla-Noarbe}, {Smith},
  {Soares-Santos}, {Suchyta}, {Swanson}, {Tarle}, {To}, {Varga}, \&
  {Walker}}]{2022ApJS..258...41B}
---. 2022, \apjs, 258, 41, \dodoi{10.3847/1538-4365/ac3914}

\bibitem[{{Bolin} {et~al.}(2023){Bolin}, {Fremling}, {Morbidelli}, {Noll}, {van
  Roestel}, {Deibert}, {Delbo}, {Gimeno}, {Heo}, {Lisse}, {Seccull}, \&
  {Suh}}]{Bolin2023}
{Bolin}, B.~T., {Fremling}, C., {Morbidelli}, A., {et~al.} 2023, \mnras, 521,
  L29, \dodoi{10.1093/mnrasl/slad018}

\bibitem[{{{\'C}uk} {et~al.}(2012){{\'C}uk}, {Hamilton}, \&
  {Holman}}]{2012MNRAS.426.3051C}
{{\'C}uk}, M., {Hamilton}, D.~P., \& {Holman}, M.~J. 2012, \mnras, 426, 3051,
  \dodoi{10.1111/j.1365-2966.2012.21964.x}

\bibitem[{Darling(1957)}]{10.2307/2237048}
Darling, D.~A. 1957, The Annals of Mathematical Statistics, 28, 823.
\newblock \url{http://www.jstor.org/stable/2237048}

\bibitem[{{DeMeo} \& {Carry}(2014)}]{2014Natur.505..629D}
{DeMeo}, F.~E., \& {Carry}, B. 2014, \nat, 505, 629,
  \dodoi{10.1038/nature12908}

\bibitem[{{Ferland} {et~al.}(2013){Ferland}, {Porter}, {van Hoof}, {Williams},
  {Abel}, {Lykins}, {Shaw}, {Henney}, \& {Stancil}}]{2013RMxAA..49..137F}
{Ferland}, G.~J., {Porter}, R.~L., {van Hoof}, P.~A.~M., {et~al.} 2013, \rmxaa,
  49, 137.
\newblock \doarXiv{1302.4485}

\bibitem[{{Fernandez} \& {Ip}(1984)}]{1984Icar...58..109F}
{Fernandez}, J.~A., \& {Ip}, W.~H. 1984, \icarus, 58, 109,
  \dodoi{10.1016/0019-1035(84)90101-5}

\bibitem[{{Gomes} \& {Nesvorn{\'y}}(2016)}]{2016A&A...592A.146G}
{Gomes}, R., \& {Nesvorn{\'y}}, D. 2016, \aap, 592, A146,
  \dodoi{10.1051/0004-6361/201527757}

\bibitem[{{Hahn} \& {Malhotra}(1999)}]{1999AJ....117.3041H}
{Hahn}, J.~M., \& {Malhotra}, R. 1999, \aj, 117, 3041, \dodoi{10.1086/300891}

\bibitem[{{Hainaut} {et~al.}(2012){Hainaut}, {Boehnhardt}, \&
  {Protopapa}}]{2012A&A...546A.115H}
{Hainaut}, O.~R., {Boehnhardt}, H., \& {Protopapa}, S. 2012, \aap, 546, A115,
  \dodoi{10.1051/0004-6361/201219566}

\bibitem[{{Hainaut} \& {Delsanti}(2002)}]{2002A&A...389..641H}
{Hainaut}, O.~R., \& {Delsanti}, A.~C. 2002, \aap, 389, 641,
  \dodoi{10.1051/0004-6361:20020431}

\bibitem[{{Hasegawa} {et~al.}(2021){Hasegawa}, {Marsset}, {DeMeo}, {Bus},
  {Geem}, {Ishiguro}, {Im}, {Kuroda}, \& {Vernazza}}]{2021ApJ...916L...6H}
{Hasegawa}, S., {Marsset}, M., {DeMeo}, F.~E., {et~al.} 2021, \apjl, 916, L6,
  \dodoi{10.3847/2041-8213/ac0f05}

\bibitem[{{Horner} \& {Lykawka}(2010)}]{2010MNRAS.402...13H}
{Horner}, J., \& {Lykawka}, P.~S. 2010, \mnras, 402, 13,
  \dodoi{10.1111/j.1365-2966.2009.15702.x}

\bibitem[{{Jewitt}(2018)}]{2018AJ....155...56J}
{Jewitt}, D. 2018, \aj, 155, 56, \dodoi{10.3847/1538-3881/aaa1a4}

\bibitem[{{Jewitt} \& {Meech}(1986)}]{1986ApJ...310..937J}
{Jewitt}, D., \& {Meech}, K.~J. 1986, \apj, 310, 937, \dodoi{10.1086/164745}

\bibitem[{{Jewitt}(2002)}]{2002AJ....123.1039J}
{Jewitt}, D.~C. 2002, \aj, 123, 1039, \dodoi{10.1086/338692}

\bibitem[{Kirch(2008)}]{ref1}
Kirch, W., ed. 2008, Pearson's Correlation Coefficient (Dordrecht: Springer
  Netherlands), 1090--1091, \dodoi{10.1007/978-1-4020-5614-7_2569}

\bibitem[{{Kortenkamp} {et~al.}(2004){Kortenkamp}, {Malhotra}, \&
  {Michtchenko}}]{2004Icar..167..347K}
{Kortenkamp}, S.~J., {Malhotra}, R., \& {Michtchenko}, T. 2004, \icarus, 167,
  347, \dodoi{10.1016/j.icarus.2003.09.021}

\bibitem[{{Lacerda} {et~al.}(2014){Lacerda}, {Fornasier}, {Lellouch}, {Kiss},
  {Vilenius}, {Santos-Sanz}, {Rengel}, {M{\"u}ller}, {Stansberry}, {Duffard},
  {Delsanti}, \& {Guilbert-Lepoutre}}]{2014ApJ...793L...2L}
{Lacerda}, P., {Fornasier}, S., {Lellouch}, E., {et~al.} 2014, \apjl, 793, L2,
  \dodoi{10.1088/2041-8205/793/1/L2}

\bibitem[{{Lin} {et~al.}(2022){Lin}, {Markwardt}, {Napier}, {Adams}, \&
  {Gerdes}}]{2022RNAAS...6...79L}
{Lin}, H.-W., {Markwardt}, L., {Napier}, K.~J., {Adams}, F.~C., \& {Gerdes},
  D.~W. 2022, Research Notes of the American Astronomical Society, 6, 79,
  \dodoi{10.3847/2515-5172/ac6752}

\bibitem[{{Lin} {et~al.}(2019){Lin}, {Gerdes}, {Hamilton}, {Adams},
  {Bernstein}, {Sako}, {Bernadinelli}, {Tucker}, {Allam}, {Becker}, {Khain},
  {Markwardt}, {Franson}, {Abbott}, {Annis}, {Avila}, {Brooks}, {Carnero
  Rosell}, {Carrasco Kind}, {Cunha}, {D'Andrea}, {da Costa}, {De Vicente},
  {Doel}, {Eifler}, {Flaugher}, {Garc{\'\i}a-Bellido}, {Hollowood},
  {Honscheid}, {James}, {Kuehn}, {Kuropatkin}, {Maia}, {Marshall}, {Miquel},
  {Plazas}, {Romer}, {Sanchez}, {Scarpine}, {Sevilla-Noarbe}, {Smith}, {Smith},
  {Soares-Santos}, {Sobreira}, {Suchyta}, {Tarle}, {Walker}, \&
  {Wester}}]{2019Icar..321..426L}
{Lin}, H.~W., {Gerdes}, D.~W., {Hamilton}, S.~J., {et~al.} 2019, \icarus, 321,
  426, \dodoi{10.1016/j.icarus.2018.12.006}

\bibitem[{{Lin} {et~al.}(2021){Lin}, {Chen}, {Volk}, {Gladman}, {Murray-Clay},
  {Alexandersen}, {Bannister}, {Lawler}, {Ip}, {Lykawka}, {Kavelaars}, {Gwyn},
  \& {Petit}}]{2021Icar..36114391L}
{Lin}, H.~W., {Chen}, Y.-T., {Volk}, K., {et~al.} 2021, \icarus, 361, 114391,
  \dodoi{10.1016/j.icarus.2021.114391}

\bibitem[{Lisse \& Steckloff(2022)}]{LISSE2022114995}
Lisse, C., \& Steckloff, J. 2022, Icarus, 381, 114995,
  \dodoi{https://doi.org/10.1016/j.icarus.2022.114995}

\bibitem[{{Luu} \& {Jewitt}(1996)}]{1996AJ....112.2310L}
{Luu}, J., \& {Jewitt}, D. 1996, \aj, 112, 2310, \dodoi{10.1086/118184}

\bibitem[{{Lykawka} {et~al.}(2011){Lykawka}, {Horner}, {Jones}, \&
  {Mukai}}]{2011MNRAS.412..537L}
{Lykawka}, P.~S., {Horner}, J., {Jones}, B.~W., \& {Mukai}, T. 2011, \mnras,
  412, 537, \dodoi{10.1111/j.1365-2966.2010.17936.x}

\bibitem[{{Magnier} {et~al.}(2013){Magnier}, {Schlafly}, {Finkbeiner}, {Juric},
  {Tonry}, {Burgett}, {Chambers}, {Flewelling}, {Kaiser}, {Kudritzki},
  {Morgan}, {Price}, {Sweeney}, \& {Stubbs}}]{Magnier2013}
{Magnier}, E.~A., {Schlafly}, E., {Finkbeiner}, D., {et~al.} 2013, \apjs, 205,
  20, \dodoi{10.1088/0067-0049/205/2/20}

\bibitem[{{Malhotra}(1993)}]{1993Natur.365..819M}
{Malhotra}, R. 1993, \nat, 365, 819, \dodoi{10.1038/365819a0}

\bibitem[{{Malhotra}(1995)}]{1995AJ....110..420M}
---. 1995, \aj, 110, 420, \dodoi{10.1086/117532}

\bibitem[{{Markwardt} {et~al.}(2021){Markwardt}, {Adams}, {Gerdes}, {Lin},
  {Malhotra}, \& {Napier}}]{2021jwst.prop.2550M}
{Markwardt}, L., {Adams}, F., {Gerdes}, D., {et~al.} 2021, {The First Near-IR
  Spectroscopic Survey of Neptune Trojans}, JWST Proposal. Cycle 1, ID. \#2550

\bibitem[{{Moffat}(1969)}]{Moffat}
{Moffat}, A.~F.~J. 1969, \aap, 3, 455

\bibitem[{{Nesvorn{\'y}} {et~al.}(2018){Nesvorn{\'y}}, {Vokrouhlick{\'y}},
  {Bottke}, \& {Levison}}]{2018NatAs...2..878N}
{Nesvorn{\'y}}, D., {Vokrouhlick{\'y}}, D., {Bottke}, W.~F., \& {Levison},
  H.~F. 2018, Nature Astronomy, 2, 878, \dodoi{10.1038/s41550-018-0564-3}

\bibitem[{{Nesvorn{\'y}} {et~al.}(2013){Nesvorn{\'y}}, {Vokrouhlick{\'y}}, \&
  {Morbidelli}}]{2013ApJ...768...45N}
{Nesvorn{\'y}}, D., {Vokrouhlick{\'y}}, D., \& {Morbidelli}, A. 2013, \apj,
  768, 45, \dodoi{10.1088/0004-637X/768/1/45}

\bibitem[{{Neveu} \& {Vernazza}(2019)}]{2019ApJ...875...30N}
{Neveu}, M., \& {Vernazza}, P. 2019, \apj, 875, 30,
  \dodoi{10.3847/1538-4357/ab0d87}

\bibitem[{{Parker} {et~al.}(2013){Parker}, {Buie}, {Osip}, {Gwyn}, {Holman},
  {Borncamp}, {Spencer}, {Benecchi}, {Binzel}, {DeMeo}, {Fabbro}, {Fuentes},
  {Gay}, {Kavelaars}, {McLeod}, {Petit}, {Sheppard}, {Stern}, {Tholen},
  {Trilling}, {Ragozzine}, {Wasserman}, \& {Ice Hunters}}]{2013AJ....145...96P}
{Parker}, A.~H., {Buie}, M.~W., {Osip}, D.~J., {et~al.} 2013, \aj, 145, 96,
  \dodoi{10.1088/0004-6256/145/4/96}

\bibitem[{Pedregosa {et~al.}(2011)Pedregosa, Varoquaux, Gramfort, Michel,
  Thirion, Grisel, Blondel, Prettenhofer, Weiss, Dubourg, Vanderplas, Passos,
  Cournapeau, Brucher, Perrot, \& Duchesnay}]{scikit-learn}
Pedregosa, F., Varoquaux, G., Gramfort, A., {et~al.} 2011, Journal of Machine
  Learning Research, 12, 2825

\bibitem[{{Peixinho} {et~al.}(2015){Peixinho}, {Delsanti}, \&
  {Doressoundiram}}]{2015A&A...577A..35P}
{Peixinho}, N., {Delsanti}, A., \& {Doressoundiram}, A. 2015, \aap, 577, A35,
  \dodoi{10.1051/0004-6361/201425436}

\bibitem[{{Peixinho} {et~al.}(2012){Peixinho}, {Delsanti}, {Guilbert-Lepoutre},
  {Gafeira}, \& {Lacerda}}]{2012A&A...546A..86P}
{Peixinho}, N., {Delsanti}, A., {Guilbert-Lepoutre}, A., {Gafeira}, R., \&
  {Lacerda}, P. 2012, \aap, 546, A86, \dodoi{10.1051/0004-6361/201219057}

\bibitem[{{Pike} {et~al.}(2017{\natexlab{a}}){Pike}, {Lawler}, {Brasser},
  {Shankman}, {Alexandersen}, \& {Kavelaars}}]{2017AJ....153..127P}
{Pike}, R.~E., {Lawler}, S., {Brasser}, R., {et~al.} 2017{\natexlab{a}}, \aj,
  153, 127, \dodoi{10.3847/1538-3881/aa5be9}

\bibitem[{{Pike} {et~al.}(2017{\natexlab{b}}){Pike}, {Fraser}, {Schwamb},
  {Kavelaars}, {Marsset}, {Bannister}, {Lehner}, {Wang}, {Alexandersen},
  {Chen}, {Gladman}, {Gwyn}, {Petit}, \& {Volk}}]{2017AJ....154..101P}
{Pike}, R.~E., {Fraser}, W.~C., {Schwamb}, M.~E., {et~al.} 2017{\natexlab{b}},
  \aj, 154, 101, \dodoi{10.3847/1538-3881/aa83b1}

\bibitem[{{Roig} \& {Nesvorn{\'y}}(2015)}]{2015AJ....150..186R}
{Roig}, F., \& {Nesvorn{\'y}}, D. 2015, \aj, 150, 186,
  \dodoi{10.1088/0004-6256/150/6/186}

\bibitem[{{Schwamb} {et~al.}(2019){Schwamb}, {Fraser}, {Bannister}, {Marsset},
  {Pike}, {Kavelaars}, {Benecchi}, {Lehner}, {Wang}, {Thirouin}, {Delsanti},
  {Peixinho}, {Volk}, {Alexandersen}, {Chen}, {Gladman}, {Gwyn}, \&
  {Petit}}]{2019ApJS..243...12S}
{Schwamb}, M.~E., {Fraser}, W.~C., {Bannister}, M.~T., {et~al.} 2019, \apjs,
  243, 12, \dodoi{10.3847/1538-4365/ab2194}

\bibitem[{{Schwarz} {et~al.}(2011){Schwarz}, {Ness}, {Osborne}, {Page},
  {Evans}, {Beardmore}, {Walter}, {Helton}, {Woodward}, {Bode}, {Starrfield},
  \& {Drake}}]{2011ApJS..197...31S}
{Schwarz}, G.~J., {Ness}, J.-U., {Osborne}, J.~P., {et~al.} 2011, \apjs, 197,
  31, \dodoi{10.1088/0067-0049/197/2/31}

\bibitem[{{Seccull} {et~al.}(2018){Seccull}, {Fraser}, {Puzia}, {Brown}, \&
  {Sch{\"o}nebeck}}]{2018ApJ...855L..26S}
{Seccull}, T., {Fraser}, W.~C., {Puzia}, T.~H., {Brown}, M.~E., \&
  {Sch{\"o}nebeck}, F. 2018, \apjl, 855, L26, \dodoi{10.3847/2041-8213/aab3dc}

\bibitem[{{Sheppard}(2010)}]{2010AJ....139.1394S}
{Sheppard}, S.~S. 2010, \aj, 139, 1394, \dodoi{10.1088/0004-6256/139/4/1394}

\bibitem[{{Sheppard}(2012)}]{2012AJ....144..169S}
---. 2012, \aj, 144, 169, \dodoi{10.1088/0004-6256/144/6/169}

\bibitem[{{Sheppard} \& {Trujillo}(2006)}]{2006Sci...313..511S}
{Sheppard}, S.~S., \& {Trujillo}, C.~A. 2006, Science, 313, 511,
  \dodoi{10.1126/science.1127173}

\bibitem[{Tabeshian {et~al.}(2019)Tabeshian, Wiegert, Ye, Hui, Gao, \&
  Tan}]{Tabeshian_2019}
Tabeshian, M., Wiegert, P., Ye, Q., {et~al.} 2019, The Astronomical Journal,
  158, 30, \dodoi{10.3847/1538-3881/ab245d}

\bibitem[{{Tonry} {et~al.}(2012){Tonry}, {Stubbs}, {Lykke}, {Doherty},
  {Shivvers}, {Burgett}, {Chambers}, {Hodapp}, {Kaiser}, {Kudritzki},
  {Magnier}, {Morgan}, {Price}, \& {Wainscoat}}]{Tonry2012}
{Tonry}, J.~L., {Stubbs}, C.~W., {Lykke}, K.~R., {et~al.} 2012, \apj, 750, 99,
  \dodoi{10.1088/0004-637X/750/2/99}

\bibitem[{{Wong} \& {Brown}(2017)}]{2017AJ....153..145W}
{Wong}, I., \& {Brown}, M.~E. 2017, \aj, 153, 145,
  \dodoi{10.3847/1538-3881/aa60c3}

\end{thebibliography}
\bibliographystyle{aasjournal}



\end{document}